# Presence Factor-Oriented Blog Summarization

Rosy Madaan
Department of Computer Science & Engineering
Echelon Institute of Technology
Faridabad, India
madaan.rosy@gmail.com

A. K. Sharma
Department of Computer Engineering
Y.M.C.A. University of Science & Technology
Faridabad, India
ashokkale2@rediffmail.com

Ashutosh Dixit
Department of Computer Engineering
Y.M.C.A. University of Science & Technology
Faridabad, India
dixit_ashutosh@rediffmail.com

*Abstract*— The research that has been carried out on blogs focused on blog posts only, ignoring the title of the blog page. Also, in summarization only a set of representative sentences are extracted. Some analysis has been done and it has been found that the blog post contains the content that is likely to be related to the topic of the blog post. Thus, proposed system of summarization makes use of title contained in a blog page. The approach makes use of the Presence factor that indicates the presence of each term of the title in each sentence of the blog post. This is a key feature because it considers those sentences as more relevant for summarization that contain each of the term present in the title. The system has been implemented and evaluated experimentally. The system has shown promising results.

**Keywords- Search Engine, crawler, WWW, blog, summarization**

## I. MOTIVATION

Search engines [2,3] provide users a way to find the required information from the World Wide Web (WWW) [1,4] in an organized fashion. For this purpose, a search engine generally uses special component called crawler [2,3] that traverse the entire Web and downloads Web pages. After downloading, these pages are then stored in repository and are passed to another program called indexer for indexing. Indexer maintains an index that maps keywords to the documents in which they appear. The index is then searched for the answering user's query and the results are returned to the users.

Blogosphere, the richest sources of information, is a part of the WWW consisting of weblogs or blogs [5,6] that are consists of dated entries typically listed in reverse chronological order on a single page. Blogs are usually personal journals maintained on the WWW that are used for writing on a variety of topics and allow the users to share their opinions and emotions. A blog is just like a web page consisting of a set of paragraphs. It contains links to several other blog pages or to other sections of the same page. The blogs are often written to be read by many others. Since, a blog page is written by either a single person or a small group, so a consistent style of writing is used across the whole text. These are the web pages that are frequently modified and are sorted in order of the date on which they have been written. Blogs can be classified according to their purpose: personal blogs (documenting one's life), issues blogs (expressing opinions, writing comments, debating current events) and topical blogs (serve as community forums allowing users share their ideas with each other). Blogs have a wide coverage. They cover most of the topics like education, entertainment, sports, music, health, business, agriculture etc. Some of the popular blog sites are technorati, wordpress, blogger, livejournal, typepad, travelpod etc.

Broadly, blogs contains the relevant in depth information, opinions and emotions of the users on particular topics. It shows that the information on blogs is very much relevant as compared to the remaining part of WWW. In order to increase the coverage of continuously growing size of the blogosphere, blog search engines continue to crawl the Web to discover, identify and index blogs. Because of the huge content available in blogosphere, the blog search engines are required to summarize the blog pages so that only relevant information can be preserved by them before indexing. Therefore, in this paper, a novel technique to summarize the blog pages has been proposed that summarize the blog pages by taking the advantage of the title in the blog pages.

The paper has been organized as follows: section2 describes the current research that has been carried out in this area; section 3 describes the proposed blog summarization system that ranks the sentences in the blogs by using title of the blog; section 4 shows the experimental work and compares the performance of the proposed system with the other research works that has been carried out in the area; last section concludes the proposed work.

## II. RELATED WORK

Since blogosphere has much relevant information and as its content and size is growing at very rapid rate, it has become the area of focus of research in recent years. Also as an emerging area, very few studies have been reported in the area of blog summarization.

Xiaodan Song et. al. in [7] proposed a system that summarizes the opinions in the massive and complex blogosphere by finding the most influential blogs with highly innovative opinions. In this work, blog networks are found where nodes represent the blogs that discuss this query and edges represent the links among blogs embedded in entries. After retrieving the blog network, the top blogs are ranked using InfluenceRank algorithm, in which blogs are ranked by how important they are to other blogs as well as the novelty of the information they contribute to the network.

To develop an effective opinion summarization approach, Shamima Mithun et. Al. in [8], have targeted to resolve specifically Question Irrelevancy and Discourse Incoherency problems which have been found to be the most frequently occurring problems for opinion summarization. To address these problems, a hybrid approach has been used by combining text schema and rhetorical relations to exploit intra-sentential rhetorical relations.

Beaux Sharifi [9], have developed an algorithm that takes a trending phrase or any phrase specified by a user, collects a large number of posts containing the phrase, and provides an automatically created summary of the posts related to the term.

In [10], Shuang Sun et. al. re-formalized the blog post summarization problem as a sentence extraction and sentence ranking problem. Three fast features, important sentences, blog tags and blog comments, have been discussed in order to calculate salience scores of representative words. An average-summation-based sentence selection method called ASS has been used to select sentences based on the salience scores of content words in sentences.

Aixin Sun et. al. in [11], extracted representative sentences from a blog post that best represent the topics discussed among its comments. The proposed solution first derives representative words from comments and then selects sentences containing representative words.

A critical review on the available literature shows the following shortcomings in the area of blog summarization:
1. The study shows that the existing systems are very complex in nature in the sense that they are using blog networks [7] some fast features, which are tough to find and thus consumes a lot of time.
2. The work in [11] made use of only the comments in the blogs and left the other features or characteristics of blog pages. Moreover, all the comments in the blog pages may or may not be valid. This can affect the performance of summarization system.

Therefore, in this paper, a blog summarization system has been proposed that best uses the most important characteristic of a blog page i.e. title of the blog.

PROPOSED SYSTEM

As discussed earlier, blogs are usually personal journals maintained on the WWW that are used for writing on a variety of topics and allow the users to share their opinions and emotions. A blog is just like a web page consisting of a set of paragraphs. Thus, it has the following main inherent parts:
- Blog Title
- Blog Post
- Visitor's Comments

Each blog page contain title related to the blog post, blog post mainly contains the opinion or emotions related to the title and visitor's comments are the comments given by blog visitors. However, the current research highly ignores the role of title of the blog in finding out the blog summary. Since, blog title is integral part of a blog page and can play important role in finding out summary of a blog page, therefore, in this research, the problem of blog summarization has been tackled by using the blog title. This work can be used in many areas such as blog search, blog presentation, reader feedback, marketing research and others.

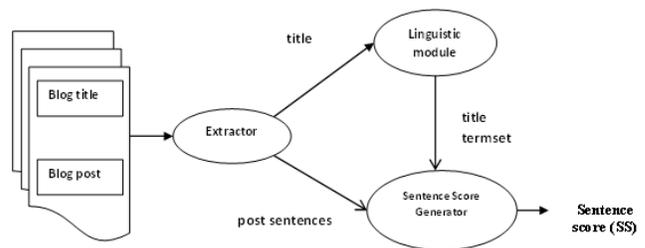

Figure 1. Proposed System for Blog Summarization

Given a blog page consisting of mainly blog Title and the blog post, the proposed system works in the following manner (see Fig. 1):
Step1: Extractor module extracts the title of the blog contained in blog page and blog post sentences on the blog page. A set is created namely a sentence set .
Step2: Blog title is given as the input to Linguistic module that performs lemmatization, stemming, normalization and stopword removal and generates title termset.
Step3: Title termset and post sentences are provided to the Sentence Score Generator module that considers all the termsets and assigns the sentence score to each sentence and sorts them according to their scores.
Step4: The sentences in the blog post are arranged according to their scores are further used to find the blog summary.

The main component of the proposed technique for blog summarization works as follows:

**Extractor:** In order to find out summary of a blog page, Extractor module extracts title from the blog page and post sentences that have been given by the visitors on the blog page.

This module is responsible for separating the sentences that exists in the blog post and generates a set of sentences S.

$$S = \{S1, S2, S3.......Sn\} \quad (i)$$

where S is the set of all sentences in a blog post and the sentences are referred to as S1, S2, S3…….Sn respectively in the order in which they appear.

**Linguistic Module:** The main task of the linguistic module is to perform the following functions:
- *Stopword Removal* process will remove the stopwords like a, the, of, for, in etc. from each sentence existing in S.
- *Lemmatization* process will perform the lemmatization. For example reducing cars to car and also automobile.
- *Normalization* process will normalize the sentence.
- *Stemming* process uses Porter's stemmer. For example reducing natural to nature.

The linguistic module will take the title and post sentences as the input and generates the title termset containing all the terms in the title and it is represented by T.

$$T = \{t1, t2, t3…….., tn\} \quad (iii)$$

where ti is the ith term in the title termset T.

**Sentence Score Generator:** This module takes title termset and post sentences as its input and computes the sentence score (SS) for each sentence.

At first, SS generator considers the title termset and sentences and generates a matrix in which each row consists of a term from the title termset and each column consist of a sentence (Si). Each entry in the matrix is represented by TSM[i, j] i.e. the frequency of term Ti of blog title in the sentence Si of the blog post (see Fig. 2).

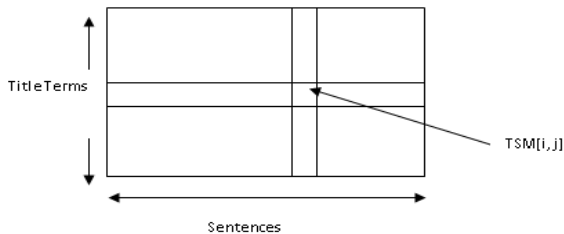

Figure 2. TSM[i,j] i.e. Title-Sentence matrix

Along with TSM [i, j], a matrix called Presence Factor Matrix PFM [i, j] is also maintained that contains title terms as rows and sentences as the column. Each entry in the matrix represented by PFM [i, j] contains presence of term (in the title) in a sentence (Si). The presence of each term in the sentence Si is represented by '1' and the absence of each term in the sentence Si is represented by '0'.

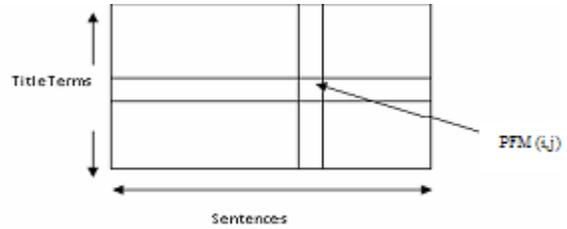

Figure 3. PFM[i,j] i.e. Presence Factor Matrix

After generation of TSM and PFM, the SS Generator computes the Sentence Score by using the following formula

$$SS(Sj) = \sum_{i=1}^{n} TSM(i, j) * PFM(i, j) \quad (viii)$$

Where
*SS (Sj)* is Sentence Score for jth sentence,
*TSM(i, j)* is the frequency of ith term of blog title in jth sentence,
*PFM(i, j)* is the presence of ith term of blog title in jth sentence. The presence is represented by '1' and absence is represented by '0'.

Maintenance of the Presence factor matrix is a key feature used in the approach. This feature ensures that the proposed technique assigns greater score to those sentences, for summary that consists of each term present in the title termset. Also, a higher score is assigned to those sentences in which more number of terms existing in the title termset is present.

Keeping this in mind, SS is computed for all the sentences in set S and thereafter, the sentences are ranked according to their Sentence Score. The top k sentences are selected as the Blog Summary. The next section discusses the experimental evaluation that justifies the proposed mechanism.

### III. EXPERIMENTAL EVALUATION

The work proposed in this paper is responsible for finding out the summary of the blog by using title in the blog page. For this purpose, the title and body sections are separated and on each section a linguistic module is applied. For each sentence in the body, the frequency of each term present in the title is evaluated. Finally, the Sentence Score is computed and the summary is presented to the user.

Several experiments have been conducted to evaluate the performance of the proposed work. The proposed mechanism has been implemented on .Net framework using - Intel Core i3, 3.2 GB RAM, 320 GB HDD. About seventy six blogs were collected from different blog sources and the proposed technique was then applied on those blog pages. Very promising results have been found. The proposed work is then compared with several online summarization tools as well as with the manual summary. It is not possible to show the summaries of all the blog pages here, thus in order to show the performance of proposed system, an example of blog page from *UandItalk* i.e. *http://uanditalk.blogspot.in/* is taken. Fig. 3 shows the summary of above mentioned blog page by the proposed system.

The analysis of the summaries generated by the proposed technique is very difficult. It needs the involvement of human summarizers as well as the systems available for automatic summarization of the blog pages. The following four online summarizers were used to find out the summaries of the blog pages.

    a.  www.freesummarizer.com
    b.  www.smmry.com
    c.  www.textcompactor.com
    d.  www.tools4noobs.com

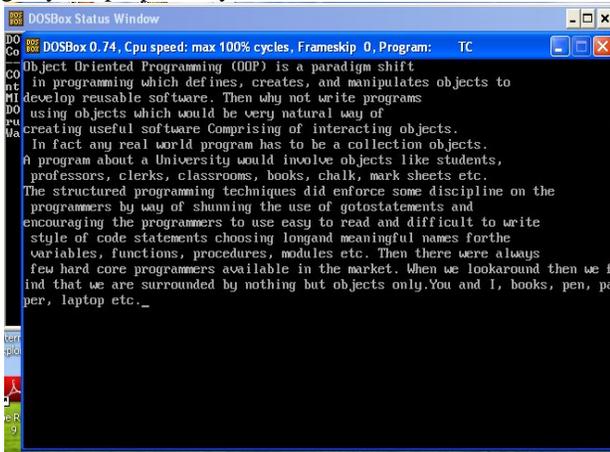

Figure 3. Summary of blog page from *http://uanditalk.blogspot.in/*

For convenience, the title of the blog, its content and the summary (by proposed system) has been shown in tabular form (see Table 1).

**Table1. Blog Title, blog content, comments and the summary (by proposed system)of** *http://uanditalk.blogspot.in/*

| Blog Title | Object Oriented Programming |
|---|---|
| **Blog Content (http://uanditalk.blogspot.in/2012/11/object-oriented-programming.html#comment-form)** | The programming process has evolved through many phases. The journey started with programmers who would write programs which somehow worked without giving any importance to readability of the program. The languages like FORTRAN and BASIC neither enforced any discipline nor were the programmers trained to write user centric programs. The result was unstructured understandable programs susceptible to bugs acting as time bombs like Y2K problem. The major problem was that the programs were not maintainable. <br> The structured programming techniques did enforce some discipline on the programmers by way of shunning the use of 'goto' statements and encouraging the programmers to use 'easy to read and difficult to write' style of code statements choosing long and meaningful names for the variables, functions, procedures, modules etc. The major emphasis was to use block structures in the program. For instance any code enclosed between a pair of curly braces and the pair of begin – end was termed as a block. Pascal and 'C' supported blocks like compound statements, loops, functions, procedures, files etc. This technique worked good for hard core programmers who were able to write large and complex programs using structured programming techniques. Unix operating system was written using 'C'. Then there were always few hard core programmers available in the market. <br> Nevertheless, the importance was given to procedures as to how to solve the problem at hand. The algorithm development consumed more time of the developer and least importance was given to the data. So, a program was virtually a collection of decomposed components  interacting functions or modules exchanging data and data structures among them. Such data, especially the global data, was vulnerable to inadvertent corruption by the fellow programmers. <br> The remedy to the above mentioned drawbacks is that we put more emphasis on data and |

|  | try to create reusable software components. The reusable components can further be combined to get bigger and more powerful software.<br>For example, in our day to day life we compose bigger objects from smaller objects. The desktop computer is made up many smaller objects like: mother board, RAM, HDD, SMPS, mouse, keyboard etc. We use the services offered by these objects and never bother as to how they work or who made them?<br>When we look around then we find that we are surrounded by nothing but objects only. You & I, books, pen, paper, laptop etc. In fact any real world program has to be a collection objects. A program about a University would involve objects like students, professors, clerks, class rooms, books, chalk, mark sheets etc. Then why not write programs using objects which would be very natural way of creating useful software Comprising of interacting objects.<br>Object Oriented Programming (OOP) is a paradigm shift in programming which defines, creates, and manipulates objects to develop reusable software. C++ is an imperative language developed to support OOP components and features like: classes, objects, Abstraction, encapsulation, inheritance, polymorphism etc.<br>This technique worked good for hard core programmers who were able to write large and complex programs using structured programming techniques.<br>A. K. Sharma |
|---|---|
| **Summary according to the proposed approach** | Object Oriented Programming (OOP) is a paradigm shift in programming which defines, creates, and manipulates objects to develop reusable software. Then why not write programs using objects which would be very natural way of creating useful software Comprising of interacting objects. In fact any real world program has to be a collection objects. A program about a University would involve objects like students, professors, clerks, class rooms, books, chalk, mark sheets etc. The structured programming techniques did enforce some discipline on the programmers by way of shunning the use of 'goto' statements and encouraging the programmers to use 'easy to read and difficult to write' style of code statements choosing long and meaningful names for the variables, functions, procedures, modules etc. Then there were always few hard core programmers available in the market. When we look around then we find that we are surrounded by nothing but objects only. You & I, books, pen, paper, laptop etc. |

Table 1 shows the summary of the blog page (*http://uanditalk.blogspot.in/*) by using the proposed summarization system.

**Table 2. Summaries generated by the proposed approach, online summarizers and human summarizer**

| **Summary produced from online tool (www.freesummarizer.com)** | The journey started with programmers who would write programs which somehow worked without giving any importance to readability of the program.<br>The structured programming techniques did enforce some discipline on the programmers by way of shunning the use of 'goto' statements and encouraging the programmers to use 'easy to read and difficult to write' style of code statements.<br>This technique worked good for hard core programmers who were able to write large and complex programs using structured programming techniques.<br>Then why not write programs using objects which would be very natural way of creating useful software comprising of interacting objects.<br>Object Oriented Programming (OOP) is a paradigm shift in programming which defines, creates, and manipulates objects to develop reusable software. |
|---|---|
| **Summary produced from online tool (www.smmry.com)** | The journey started with programmers who would write programs which somehow worked without giving any importance to readability of the program.<br>The languages like FORTRAN and BASIC neither enforced any discipline nor were the programmers trained to write user centric programs.<br>The structured programming techniques did enforce some discipline on the programmers by way of shunning the use of 'goto' statements and encouraging the programmers to use 'easy to read and difficult to write' style of code statements i.e.<br>This technique worked good for hard core programmers who were able to write large |
| | and complex programs using structured programming techniques.<br>Then why not write programs using objects which would be very natural way of creating useful software comprising of interacting objects.<br>Object Oriented Programming (OOP) is a paradigm shift in programming which defines, creates, and manipulates objects to develop reusable software. |
| **Summary produced from online tool ([www.textcompactor.com](www.textcompactor.com))** | The major problem was that the programs were not maintainable.<br>The structured programming techniques did enforce some discipline on the programmers by way of shunning the use of 'goto' statements and encouraging the programmers to use 'easy to read and difficult to write' style of code statements.<br>This technique worked good for hard core programmers who were able to write large and complex programs using structured programming techniques. The reusable components can further be combined to get bigger and more powerful software.<br>For example, in our day to day life we compose bigger objects from smaller objects.<br>Then why not write programs using objects which would be very natural way of creating useful software comprising of interacting objects.<br>Object Oriented Programming (OOP) is a paradigm shift in programming which defines, creates, and manipulates objects to develop reusable software. |
| **Summary produced from online tool ([www.tools4noobs.com](www.tools4noobs.com))** | I totally agree with your thought about object oriented programming about the ease of programmers. It facilitated the programmers a lot. Also provided with a different view point of programming. Yes, the object oriented technique provided the structured programming techniques encouraged the programmers to use long and meaningful names for the variables, functions, procedures, modules etc.<br>The structured programming techniques did enforce some discipline on the programmers by way of shunning the use of 'goto' statements and encouraging the programmers to use 'easy to read and difficult to write' style of code statements.<br>Choosing long and meaningful names for the variables, functions, procedures, modules etc.<br>Object Oriented Programming (OOP) is a paradigm shift in programming which defines, creates, and manipulates objects to develop reusable software.<br>Then why not write programs using objects which would be very natural way of creating useful software comprising of interacting objects.<br>This technique worked good for hard core programmers who were able to write large and complex programs using structured programming techniques. |
| **Manual Summary** | The programming process has evolved through many phases.<br>The languages like FORTRAN and BASIC neither enforced any discipline nor were the programmers trained to write user centric programs.<br>The remedy to the drawbacks is that we put more emphasis on data and try to create reusable software components.<br>Then why not write programs using objects which would be very natural way of creating useful software Comprising of interacting objects.<br>Object Oriented Programming (OOP) is a paradigm shift in programming which defines, creates, and manipulates objects to develop reusable software. |
| **Summary according to the proposed approach** | Object Oriented Programming (OOP) is a paradigm shift in programming which defines, creates, and manipulates objects to develop reusable software. Then why not write programs using objects which would be very natural way of creating useful software Comprising of interacting objects. In fact any real world program has to be a collection objects. A program about a University would involve objects like students, professors, clerks, class rooms, books, chalk, mark sheets etc. The structured programming techniques did enforce some discipline on the programmers by way of shunning the use of 'goto' statements and encouraging the programmers to use 'easy to read and difficult to write' style of code statements choosing long and meaningful names for the variables, functions, procedures, modules etc. Then there were always few hard core programmers available in the market. When we look around then we find that we are surrounded by nothing but objects only. You & I, books, pen, paper, laptop etc. |

Moreover, a human summarizer is also taken into consideration to find out the summary for the blog page. The summaries generated as the result from all these is shown in Table 2 i.e. Table 2 shows the summaries generated by the proposed system, the online summarization tools and by a human summarizer.

The summary generated by the human summarizer is considered as the *model* summary. For performance evaluation of the summaries generated above, we calculate *precision* and *recall* of each of the above summaries which are defined as follows:

1. *Precision* is defined as a fraction of number of common sentences given by *Ncommon* (sentences appearing in both, the summary under consideration and model summary) over the total number of sentences appearing in the summary under consideration i.e. *Nsumm*.
   *Precision* is given by $P = Ncommon/ Nsumm$.

2. *Recall* is defined as a fraction of number of common sentences given by *Ncommon* (sentences appearing in both, the summary under consideration and model summary) over the total number of sentences appearing in the model summary i.e. *Nmsumm*.
   *Recall* is given by $R = Ncommon/ Nmsumm$.

*Table 3. Precision and Recall* of the summaries

| Approach/tool used | Precision(%) | Recall(%) |
|---|---|---|
| Summary according to the proposed approach | 85.7 | 85.7 |
| Summary produced from online tool (www.freesummarizer.com) | 28.5 | 28.5 |
| Summary produced from online tool (www.smmry.com) | 57.1 | 57.1 |
| Summary produced from online tool (www.textcompactor.com) | 28.5 | 28.5 |
| Summary produced from online tool (www.tools4noobs.com) | 42.8 | 42.8 |

*Precision* and r*ecall* of the summaries is shown in Table 3 and the graph for the same is plotted in Fig. 4.

A careful analysis of the summary from the proposed system, the online summarizers and manual summarizer shows that the proposed method works well and produces a summary of good quality with high *precision* and *recall*.

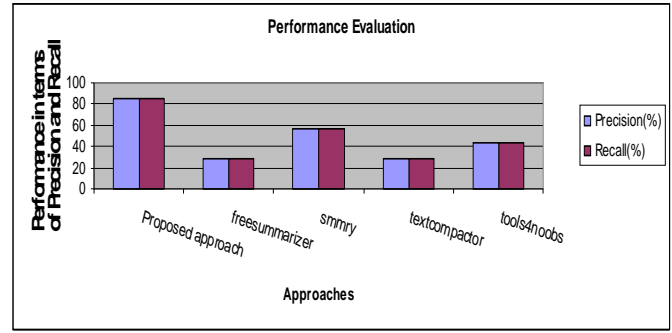

Also, it has been found that many of the sentences that have been selected by the online summarizers for summary, don't contain terms present in the title termset. Thus, it is concluded that the proposed technique of summarization works very well in summarization of blogs wherein the blog title is provided and uses Presence factor as a major criterion for selection of sentences that best represent the blog.

## IV. CONCLUSION

Much existing research that has been carried out in this area ignores the title of the blog page. We define the problem of blog post summarization by considering the title of the blog page and presence of terms of the title termset i.e. our proposed solution measures Sentence score of each sentence of the blog post by using the terms contained in the title for generating a summary of good quality. Using the PF, the quality of the generated summary has improved.

Then the proposed approach has been applied on a number of blog pages and the summaries generated have then been compared with those generated from the online systems by using model summary and *precision* and *recall* as the criterion for performance evaluation. It has been found that the system works very well for extracting relevant content from the blog posts by producing summary of good quality.